\RequirePackage{fix-cm}
\documentclass[twocolumn]{svjour3}          
\smartqed  
\usepackage{amsmath}
\usepackage{wrapfig}
\usepackage[noadjust]{cite}
\usepackage{adjustbox}
\usepackage{amssymb}
\usepackage{graphicx}
\usepackage{epstopdf}
\usepackage{epsfig}
\usepackage{hyperref}
\usepackage{newcent}
\usepackage[lined,boxed,commentsnumbered]{algorithm2e}
\usepackage{multirow}
\usepackage{subfig}
\usepackage{latexsym}
\journalname{Wireless Networks}
\begin{document}

\title{Greedy-Knapsack Algorithm for Optimal Downlink Resource Allocation in LTE Networks}


\author{ Nasim Ferdosian \and Mohamed Othman \and Borhanuddin Mohd Ali \and Kweh Yeah Lun}


\institute{N. Ferdosian \and M. Othman \and K. Y. Lun \at
              Communication Technology and Network Department, Universiti Putra Malaysia, 43400 UPM, Serdang, Selangor D.E., Malaysia \\
              Tel.: +603 8947 1707\\
              Fax: +603 8946 6576\\
              \email{n.ferdosian@gmail.com}, {mothman@upm.edu.my}
           \and
           B. M. Ali \at
           Computer and Communication Systems Department, Universiti Putra Malaysia, Serdang, Malaysia \\
}

\maketitle
\sloppy
\begin{abstract}
The  Long  Term  Evolution  (LTE)  as  a  mobile  broadband technology supports a wide domain of communication services with different requirements. Therefore, scheduling of all flows from various applications in overload states in which the requested amount of bandwidth exceeds the limited available spectrum resources is a challenging issue. Accordingly, in this paper, a greedy algorithm is presented to evaluate user candidates which are waiting for scheduling and select an optimal set of the users to maximize system performance, without exceeding available bandwidth capacity. The greedy-knapsack algorithm is defined as an optimal solution to the resource allocation problem, formulated based on the fractional knapsack problem. A compromise between throughput and QoS provisioning is obtained by proposing a class-based ranking function, which is a combination of throughput and QoS related parameters defined for each application. The simulation results show that the proposed method provides high performance in terms of throughput, loss and delay for different classes of QoS over the existing ones, especially under overload traffic.

\end{abstract}

\section{Introduction}
\noindent  The expected increase in the growth rate of mobile data is driving the evolution of mobile communication technologies. The LTE specification, which has been introduced by the 3GPP is one of the solutions for the increasing appeal of packet-based mobile broadband systems \cite{r1}. It is intended as a system to be able to provide a significant improvement in throughput over the preceding mobile standards (e.g. GSM, UMTS, HSPA) and support multiple  classes  of Quality of Service (QoS).

With the aim to efficiently support the current high variety of services, the efficient use of limited share bandwidth is essential. The purpose of effective scheduling strategies is crucial to meet the LTE targets, inasmuch as selecting an appropriate scheduling scheme is not standardized by the 3GPP specification for LTE \cite{r2}, but it is left to the vendors as an implementation decision to adaptively configure and implement an appropriate algorithm according to the desired concerns of the system \cite{r3}. However, typically, it is impossible to accomplish all intended goals at the same time \cite{r4}. Each factor can be supplied always at the cost of reducing another one. For example, scheduling algorithms aiming to optimize only spectral efficiency are unsuitable for dealing with guaranteed bit rate traffic \cite{r5}. In this sense, the main challenge is designing an allocation strategy to create a trade-off among the system performance factors. Therefore throughput-aware approaches must be used with QoS-aware strategies to provide a good balance between multi-QoS provisioning to support mixes of real-time and non-real-time traffic and system throughput maximization \cite{r3,r6}.

It is important to note that, despite the network-wide control mechanisms to mitigate traffic overload \cite{r7,r8,r9}, mobile data content overwhelms the available bandwidth for each node in many peak traffic times \cite{r10}. Based on this premise, it is clear that the overload state is an inevitable issue in LTE mobile networks, and that the proposed scheduling algorithms must also be resistant to the unexpected traffic overload patterns.

Accordingly, in this paper the design of an optimal scheduling algorithm for LTE downlink by considering the QoS requirements specified for each application is addressed. We intend to clarify how the demand of the bearers in the application architecture should be accommodated by assigning the available radio resources such that, the application requirements and resource constraints in the network are satisfied without sacrificing the system throughput.

The strategy is based on the concept of optimization problems in which the resource allocation problem is formulated as a knapsack problem. This allows for quick and accurate solutions to find nearly optimal allocation decisions. Thus, we propose a greedy heuristic approach as an efficient solution to this problem. In addition, by applying a ranking model to the bearers, the relative merits of various bearers demanding resource allocation are quantified. When this greedy-knapsack approach is applied in time-domain scheduling, it is shown to be effective in improving system performance and reliability while the network experiences a mix of normal and overload traffic. The performance evaluation is conducted in terms of average system throughput, delay and packet loss for Guaranteed Bit Rate (GBR) and Non-Guaranteed Bit Rate (Non-GBR) classes of services. In addition to the average throughput, fairness among VoIP bearers is evaluated as another effect of throughput-aware decision making for scheduling.

This paper is organized into six sections including this introductory section. Section 2 presents the main motivation for optimal greedy scheduling. It gives an overview of scheduling issues in LTE networks as well as different solutions proposed to provide efficient resource allocation among User Equipments (UEs) and cites related works. Section 3 describes the system model and provides the details of our proposed greedy-knapsack scheduling algorithm. The simulation process, simulation parameters and performance metrics are described in Section 4, while Section 5 discusses the performance of the greedy-knapsack algorithm in respect of the simulation results. Section 6 concludes the overall research study and outlines recommendations for future work.

\section{Background and Related Research}
\noindent
This section addresses three issues that are dealt with in this study, namely, LTE performance targets, optimized scheduling and reliable scheduling in overload states. It also reviews various solutions proposed in LTE literature  to provide efficient, optimized and reliable downlink resource allocation among bearers from different classes of applications in LTE networks.

It is worth noting that the responsibility for providing scheduling performance targets is up to the implementation of the eNodeB residing in the MAC layer \cite{r11}. The eNodeB assigns each active user a fraction of the total system bandwidth to share available resources among them by using a multiple access technique. The LTE downlink 3GPP adopts Orthogonal Frequency Division Multiple Access (OFDMA) as an access technique to accommodate user equipments with a wide variety of QoS application requirements and in channel conditions \cite{r12}. OFDMA allows multiple access by allocating a disjoint selective collection of sub-carriers to each individual user to leverage multi-user diversity and provide high scalability and robustness \cite{r13}. The LTE radio resources are distributed in time and frequency domains. Each OFDMA frame comprises ten 1ms sub-frames in the time domain and a sub-channel of 12 consecutive same size sub-carriers that cover 180 kHz of the frequency domain. The basic resource unit for mapping sub-carriers to active users is called the Resource Block (RB). Each RB spans over a 0.5 ms time extent and one sub-channel \cite{r14}.

To decrease the complexity and increase the design flexibility, the new sophisticated radio resource management mechanisms follow a two level framework \cite{r15, r16, r17}. In these kinds of framework the resource allocation procedure is divided between the Time Domain (TD) and Frequency Domain (FD) schedulers operating in independent ways. The TD scheduler determines a priority list of users in each Transmission Time Interval among those waiting for scheduling, and the FD scheduler is in charge of physically assigning frequency resources to the selected users in the time domain. The scheduling algorithm presented in this paper focuses on the time domain to provide efficient sharing of time resources among the candidate users. This algorithm can be integrated with most of the FD schedulers in the literature to exploit the provisioned spectrum resources efficiently.

\subsection{Performance Guarantee}
\noindent
The concept of performance guarantee in network utilization can be expressed in terms of QoS and throughput provisioning. In the field of cellular communication networks, the term QoS indicates a measure of how efficiently and reliably a network can fulfil a guaranteed level of satisfaction for its diverse services from real-time to non-real-time services. As the different services are susceptible to different measures of quality, a proper level of QoS requirements needs to be guaranteed.

Several studies have conceived the concept of QoS provisioning in LTE networks \cite{r18}. A QoS-oriented scheduler for Best Effort and Constant Bit Rate traffic was introduced in \cite{r19}. Prioritization of users is done using the common well-known Blind Equal Throughput (BET) and Proportional Fair (PF) approaches in time domain. The BET and PF metrics are expressed as:

\begin{equation}
  M_{BET}=1/(R[n])
\end{equation}
and\begin{equation}
M_{PF}=(\bar{D}[n])/(R[n])
\end{equation}
where $n$ is the user index, $\bar{D}[n]$ is the wideband throughput expected for the user  $n$ through available bandwidth and $R [n]$  is the past average throughput of user $n$, which is updated at every Transmission Time Interval (TTI) \cite{r20}. The main focus of this method is on improving total throughput along with considering guaranteed bit rate measurements as the only one QoS parameter.

A flexible fairness and QoS-oriented  multi-stream scheduler is based on the two-stage (time and frequency domains) proportional fair scheduling principle described in \cite{r21} for real-time video traffic. The Multi-Stream PF (QoS-MSPF) resource allocation algorithm considers the arrival rate and head of line packet delay to be QoS constraints. Eventually, this scheduler uses a metric, which is a combined function of delay, throughput and Channel Quality Identifier (CQI) factors, as follows:
\begin{equation}
M_{QoS-MSPF} = F_{D}\times F_{CQI}\times F_{T}
\end{equation}
where $F_{D}$ is a function of QoS delay factors, $F_{CQI}$ is a function of CQI indicating channel state information of each user, and $F_{T}$ is a factor of corresponding throughput calculation. This algorithm cannot be considered to be a strong QoS provisioning scheduler because it ignores other QoS factors, such as the minimum data-rate requirements.

The authors in \cite{r22}  applied a self-optimization method in response to the active changes in network conditions and traffic over time, and proposed an Optimized-Service Aware (OSA) scheduler. To simplify the complexity of the resource allocation procedure, it has been partitioned into three separate stages - QoS classes identified classification, time domain, and frequency domain scheduling. The OSA algorithm sorts each GBR bearer according to the Head of Line (HOL) packet delay in the buffer of the related bearer, while the non-GBR bearer list is ordered according to the following priority metric:

\begin{equation}
M_{OSA}=(\bar{D}[n])/(\theta[n])W_{QoS}
\end{equation}
where $\theta[n]$ is the normalized average channel condition estimate of bearer $\emph{n}$ and $W_{QoS}$ is the QoS weight. Two created sorted candidate groups are passed through the FD scheduler to be assigned to the optimal spectrum. The FD scheduler allocates the best RB to the highest GBR priority bearer. After giving enough resources to all GBR bearers, if any RB still remains, the FD scheduler assigns them to non-GBR bearers. The OSA algorithm can be demonstrated to be unsuitable for dealing with bounded losses as another factor of QoS support.

We note that none of the aforementioned approaches offer any strict guarantees on all QoS characteristics specified  for different classes of QoS, which plays a major role in end-user satisfaction. Consequently, a generic ranking function has been proposed in \cite{r28} to provide absolute delay and loss guarantees to GBR and Non-GBR bearers. This scheme does not account for past throughput experienced by users. Research shows that a lack of throughput awareness by schedulers when a specified level of QoS is required will lead to the quality degradation of video and VoIP services even during the low traffic times of network. Consequently, there is a need for a unique resource allocation mechanism for both QoS and throughput gain to be employed in eNodeB. In this research, we propose a ranking function, which is a combination of the influential parameters that can provide QoS provisioning as well as throughput gain in one algorithm.

\subsection{Resource Allocation Optimization}
\noindent
The LTE scheduler has to supply QoS requirements of mobile communication and be adaptive to the channel quality fluctuations. Hence, the complexity of the LTE scheduling problem has been classified as a NP-hard category \cite{r23}. The scheduling job is becoming increasingly complex because it has to be performed in hard real time fashion.

Recently, to improve the performance of the proposed schemes, cross-layer formulations and several optimization tools, including game theory, have been adopted. Nevertheless, many drawbacks that limit  the  application  of  state-of-the-art  algorithms  to practical contexts still hold, mainly due to the high computational complexity and the weak scalability of the proposed  techniques,  that  often  make  real-time  solutions intractable problems, as they also require a considerable amount of feedback information across the network nodes.

In \cite{r24}, an linearized optimization model has been presented for multi-user scheduling in the context of LTE downlink. However, obtaining an exact optimal scheduling solution remains very time-consuming. The authors in \cite{r25} proposed near-optimal scheduling approaches based on the Genetic Algorithm (GA) and Simulated Annealing (SA) heuristic methods to solve the earlier formulated optimization problem of multi-user scheduling. They provided near-optimal solutions in terms of average total bit rate, without concern for the QoS requirements, specified as loss and delay constraints.

In this respect, we rely on heuristics \cite{r27} to find solutions in a reasonable amount of time with low complexity. In \cite{r28}, a knapsack optimization implementation is employed in response to the need for QoS provisioning for LTE downlink scheduling. In this method, the LTE resource allocation problem is formulated to a fractional knapsack problem by mapping their respective properties. In this method, bearers are selected for scheduling according to their overall rank. However choosing the highest rank bearer without any concern for the extent of their required resources does not lead to optimal benefit. Furthermore, other LTE performance targets also need to be considered in resource allocation.

Accordingly, the central contribution of this paper involves the introduction and analysis of a new LTE downlink resource allocation strategy with respect to a utility function to express allocation preferences and represent the optimal allocation solution as a greedy-knapsack algorithm. Although both utility models \cite{r29, r30} and fractional knapsack resource allocation formulations are available in the literature \cite{r28}, this work presents a new and significant solution to provide optimal trade-off between the performance targets.

\section{Optimal Downlink Resource Allocation Model}
\noindent
To optimize the various objectives of proportional fairness and maximize the average system throughput, subject to satisfying QoS constraints on data rate, priority, packet loss and delay, the scheduling approach presented here is a bearer-level QoS control approach. A bearer is established between the User Equipment (UE) and Packet Data Network Gateway to indicate the application data flows within the Evolved Packet System. Typically, a user may apply several applications having different QoS requirements at the same time (as a case in point, streaming a video whilst downloading a FTP file). In order to discriminate between these various services, the QoS characteristics have been standardized in 3GPP technical specifications, in nine Quality Channel Indicator (QCI) classes as listed in Table \ref {table 1}. Thus, every bearer is assigned to a distinc QCI class and associated with an individual Allocated and Retention Priority (ARP) parameter \cite{r32}. The ARP parameter has no effect on scheduling decisions as it is used for Call Admission Control to imply the importance of the bearer set up and modification request depending on the system resource availability.

\begin{table*}[htbp]
\caption{Standardized QoS characteristics in QCI classes \cite{r31}}

\begin{tabular*}{\textwidth}{lllllp{9cm}}

\hline
\hline
\scriptsize QCI & \scriptsize Bearer type              & \scriptsize Priority &\scriptsize \begin{tabular}[c]{@{}l@{}}Packet delay \\ budget (ms)\end{tabular} & \scriptsize \begin{tabular}[c]{@{}l@{}}Packet error\\ loss rate\end{tabular} & \scriptsize Example services \\ \hline
\scriptsize 1   & \scriptsize \multirow{4}{*}{GBR}     & \scriptsize 2        & \scriptsize 100                                                                 & \scriptsize $10^{-2}$                                                               & \scriptsize Conversational Voice \\ \cline{1-1} \cline{3-6}
\scriptsize 2   & \scriptsize                         & \scriptsize 4        & \scriptsize 150                                                                 & \scriptsize $10^{-3}$                                                               & \scriptsize Conversational Video (Live Streaming)  \\ \cline{1-1} \cline{3-6}
\scriptsize 3   & \scriptsize                         & \scriptsize 3        & \scriptsize 50                                                                  & \scriptsize $10^{-3}$                                                               & \scriptsize Real Time Gaming \\  \cline{1-1} \cline{3-6}
\scriptsize 4   & \scriptsize                         & \scriptsize 5        & \scriptsize 300                                                                 & \scriptsize $10^{-6}$                                                               & \scriptsize \begin{tabular}[c]{@{}l@{}}Non-Conversational Video (Buffered Streaming)\end{tabular} \\ \hline
\scriptsize 5   & \scriptsize \multirow{5}{*}{Non-GBR} & \scriptsize 1        & \scriptsize 100                                                                 & \scriptsize $10^{-6}$                                                               & \scriptsize IMS Signalling \\  \cline{1-1} \cline{3-6}
\scriptsize 6   & \scriptsize                         & \scriptsize 6        & \scriptsize 300                                                                 & \scriptsize $10^{-6}$                                                               & \scriptsize \begin{tabular}[c]{@{}l@{}}Video (buffered streaming) TCP-based  (e.g., www, e-mail, chat, ftp, p2p file \\ sharing, progressive video, etc.)\end{tabular} \\  \cline{1-1} \cline{3-6}
\scriptsize 7   & \scriptsize                         & \scriptsize 7        & \scriptsize 100                                                                 & \scriptsize $10^{-3}$                                                              & \scriptsize \begin{tabular}[c]{@{}l@{}}Voice, Video (Live Streaming) Interactive Gaming \end{tabular} \\  \cline{1-1} \cline{3-6}
\scriptsize 8   & \scriptsize                         & \scriptsize 8        & \scriptsize 300                                                                 & \scriptsize $10^{-6}$                                                               & \scriptsize \multirow{2}{*}{\begin{tabular}[c]{@{}l@{}}Video (buffered streaming) TCP-based (e.g., www, e-mail, chat, ftp, p2p file \\ sharing, progressive video, etc.)\end{tabular}} \\  \cline{1-1} \cline{3-5}
\scriptsize 9   &  \scriptsize                        & \scriptsize 9        & \scriptsize 300                                                                 & \scriptsize $10^{-6}$                                                               &
\\ \\ \hline

\end{tabular*}
\label{table 1}
\end{table*}

\subsection{Problem Formulation}
\noindent
Every QCI has a particular packet forwarding treatment according to their predefined specifics; therefore, all bearers assigned to a especial QCI must follow a common rate and scheduling policy while fulfilling the QoS class-based constraints as well as maximizing system throughput.
In this context, three optimization objectives including, maximizing system throughput, minimizing packet loss rate and minimizing delay as respectively expressed by Equations (\ref {eq 5}),(\ref {eq 6}) and (\ref {eq 7}) should be achieved. Having these three targets in mind, the QoS class-based scheduling problem is formulated as a multi-objective optimization problem defined by the following expressions.

Objective functions: \begin{equation}
\max \sum_{i=1}^{n}\sum_{rb\in RB_{i}} r_{i,rb},
\label{eq 5}
\end{equation}
\begin{equation}
\forall i\in N: \min ( l_{i} )
\label{eq 6}
\end{equation}
and\begin{equation}
\forall i\in N: \min ( d_{i} )
\label{eq 7}
\end{equation}
subject to:
\begin{equation}
\forall i\in N, k\in K : d_{i_{k}}<D_{k},
\label{eq 8}
\end{equation}
\begin{equation}
\forall i\in N, k\in K : l_{i_{k}}<L_{k},
\label{eq 9}
\end{equation}
\begin{equation}
\forall i,j\in N , i\neq j: RB_{i}\cap RB_{j}=\emptyset,
\label{eq 10}
\end{equation}
\begin{equation}
\text{case} \ k<5: \ GBR<\overline{Rsch_{i_{k}}}<MBR,
\label{eq 11}
\end{equation}

and

\begin{equation}
\begin{split}
& \text{case} \ k\geq 5: Aggregate Bit Rate_{user_{k}} < \min (AMBR_{user_{k}},
\\ & \sum_{active APN}AMBR_{APN}),
\label{eq 12}
\end{split}
\end{equation}
where $ i\in N=\{1,2,\ldots,n\} $ denotes the index of the bearer selected for scheduling. Each bearer $i$ is assigned to an individual QCI class with label $ k\in K=\{1,2,\ldots,9\} $ and transmits data over $ | RB_{i} | $ resource blocks. $  RB_{i} $ is the set of resource blocks dedicated to the bearer $i$, where $  RB_{i}\subset RB=\{1,2,\ldots,rb\} $. Let $r_{i,rb}$ be the achieved data rate by bearer $i$ over the $rb^{th}$ resource block, $D_{k}$ and $L_{k}$ are standardized packet delay budget and packet error loss rate thresholds of the corresponding QCI class $k$ respectively, $d_{i_{k}}$ and $l_{i_{k}}$ are the measured packet delay and loss for bearer $i$ from QCI class $k$ respectively, and $\overline{Rsch_{i_{k}}}$ is average data rate achieved by bearer $i$ from QCI class $k$ when scheduled.

Two main QoS constraints stated in Inequality (\ref {eq 8}) and (\ref {eq 9}) indicate that the measured values of packet delay and loss for bearer $i$ should be less than their predefined threshold values respectively. The sets of all the assigned resource blocks are disjoint as indicated in Equation (\ref {eq 10}), implying that no resource block can be granted to more than one bearer at the same time slot. In this context, we consider a cell of LTE network with $ n $ bearer queued at the buffer of eNodeB, waiting for scheduling. As can be seen from Table \ref{table 1}, there are two broad categories of QoS bearers based on their rate policy: GBR and Non-GBR bearers. Each GBR bearer intends for the target data rate to comply with the value of the GBR; however, it is not allowed to exceed the value of Maximum Bit Rate (MBR) QoS parameter. Although the MBR value is set equal to the GBR in 3GPP release 8, the feasibility of MBR adjustment to a value greater than GBR (in later 3GPP releases) leads to the greater adaptivity support for real-time applications as considered in this research and explicitly stated in Inequality (\ref {eq 11}).

Furthermore, the network operators can bound the amount of data rate provided for any subscriber by adjusting the Aggregate Maximum Bit Rate (AMBR) parameter per APN and subscriber. Then the aggregate bit rate of an actual user must be less than the minimum of the two different AMBR values and expressed by Inequality (\ref {eq 12}).

\subsection{Optimal Greedy-Knapsack Algorithm}
\noindent
The ranked list solution of the knapsack approach \cite{r28} improves the performance of the network in terms of fulfilling the QoS constraints in a single pass; however, it is unable to define the optimum resource allocation decision. On the other hand, choosing the highest ranked bearers without any concern about the extent of the required resources by each bearer does not lead to optimal benefit. Hence, there is a need for an efficient approach to the resource allocation problem in an optimal manner. The optimal allocation can be achieved by exploiting the fact that different data bearers have a different amount of demanded radio resources. Supposing each bearer $i\in N$ has a rank $\rho_{i}$ and a size $s_{i}$, the objective is to select a set proportion of data bearers with the highest value so that the total value is maximum, but not exceeding the whole capacity of the available bandwidth $B$. The corresponding fractional knapsack model of the resource allocation problem can be expressed as
\begin{equation}
\begin{split}
& \max \sum_{i=1}^{n} x_{i}\rho_{i} \\
& \text{subject to:} \\
& \forall i\in N, 0\leq x_{i} \leq 1:  \sum_{i=1}^{n} x_{i}s_{i} \leq B
\end{split}
\label{eq:13}
\end{equation}

where $x_{i}$ is the proportion of bearer $i$ selected for resource allocating. The bearer size $s_{i}$ of a given bearer is achieved by calculating the number of its required resource blocks. Further, the rank of a given bearer, indicating the importance level of the bearer for scheduling, is computed by using a ranking function, which is expressed in detail in section 3.3.

\begin{algorithm}
\SetKwInOut{Input}{input}\SetKwInOut{Output}{output}
\Input{A queue of $n$ bearers waiting for scheduling with their corresponding rank $\rho_{i}$ and size $s_{i}$, total number of available resource blocks $B$}
\Output{array $X$ of amount $x_{i}$ of each bearer to maximize total benefit}
\BlankLine
\Begin{
\For{$i\leftarrow 1$ \KwTo $n$}{
order bearers according to their ratio $(\rho_{i} /s_{i} )$ decreasingly and index them from $1$ to $n$;

$X[i] \longleftarrow 0$\;
}
$A \longleftarrow B$\;
$i\leftarrow 1$\;

\While{$A >0$  \textbf{and}  $i\leq n$}{
  \eIf{$(A-s_{i})\geq 0$}{
   $X[i] \longleftarrow 1$\;
    $A \longleftarrow A-s_{i}$\;
    $i \longleftarrow i+1$\;
   }{
     $X[i] \longleftarrow A/s_{i}$\;
     $A \longleftarrow A-(A/s_{i}*s_{i})$\;
  }
 }return $X$\;
}
\caption{Greedy-Knapsack Scheduling Algorithm}
\label{alg1}
\end{algorithm}
As pointed out by Cormen et al. in Section 16.2 of \cite{r33}, a greedy algorithm that iteratively selects an object with better ratio $\frac{\rho_{i}}{s_{i}}$ creates an optimal solution to the fractional knapsack problem aiming to maximize $\sum x_{i}\rho_{i}$. Consequently, due to the greedy-choice property of the fractional knapsack problem of resource allocation, it can be solved optimally by using the greedy-knapsack algorithm (Algorithm \ref{alg1}) which selects  bearers with better ratio $\frac{rank\  value}{required\ RBs}$. First, the bearers are sorted by their mentioned ratio in decreasing order, and then the required number of resources from the available bandwidth are obtained, one by one, until all the available resource blocks are allocated or all the queued bearers receive the needed resources. If a bearer cannot completely fit to the remaining resources, a fraction of its packets is selected to be resource allocated such that the remaining resources are finished entirely.

The optimal performance of the proposed greedy-knapsack algorithm with respect to selecting an optimal set of bearers can be mathematically analyzed by proving the following theorem.

\textbf{Theorem} \ \emph{Algorithm \ref{alg1} can guarantee finding an optimal solution for the resource allocation problem stated in (\ref{eq:13}).}

\emph{Proof} \ Let  $X=\{x_{1},\ldots,x_{n}\}$ be the output solution by Algorithm \ref{alg1}. Primarily all the bearers have been sorted in monotonically decreasing order of $\frac{\rho_{i}}{s_{i}}$ and indexed from $1$ to $n$, where $\frac{\rho_{1}}{s_{1}}\geq \frac{\rho_{2}}{s_{2}}\geq \cdots \geq \frac{\rho_{n}}{s_{n}}$.
If for all $i$, we have $x_{i}=1$ then the output solution $X$ is optimal.

Let $j$ be the index of the first bearer for which $x_{j}<1$ and $i\neq j$ and assume that $Y=\{y_{1},\ldots,y_{n}\}$ be any feasible solution satisfying the problem constraint $\sum_{i=1}^{n} y_{i}s_{i}\leq B$. According to the algorithm \ref{alg1}, since $\sum_{i=1}^{n} x_{i}s_{i} = B$, then:
\begin{equation}
 \sum_{i=1}^{n} y_{i}s_{i}\leq \sum_{i=1}^{n} x_{i}s_{i} \end{equation}
\begin{equation}
 0\leq \sum_{i=1}^{n} (y_{i} - x_{i})s_{i}
 \label{eq 15}
\end{equation}
and also according to algorithm \ref{alg1} we have:
\begin{eqnarray*}
\text{If} \ i<j \ \text{then} \ x_{i}=1 , (x_{i}-y_{i})\geq 0 \ \text{and} \ (\rho_{i} /s_{i} )\geq(\rho_{j} /s_{j}),
\\\text{therefore} \ (x_{i}-y_{i})(\rho_{i} /s_{i} )\geq (x_{i}-y_{i})(\rho_{j} /s_{j}).
\end{eqnarray*}
\begin{eqnarray*}
\text{If} \ i>j \ \text{then} \ x_{i}=0 , (x_{i}-y_{i})\leq 0 \ \text{and} \ (\rho_{i} /s_{i} )\leq(\rho_{j} /s_{j}),
 \\\text{thus} \ \ (x_{i}-y_{i})(\rho_{i} /s_{i} )\geq (x_{i}-y_{i})(\rho_{j} /s_{j} ).
\end{eqnarray*}

\noindent Let $P (Z)$ denotes the total profit of a feasible solution $Z$, then we have:\\
\begin{eqnarray*}
P(X)-P(Y) & = & \sum_{i=1}^{n}(x_{i}-y_{i})\rho_{i}
\\ & = & \sum_{i=1}^{n}(x_{i}-y_{i})s_{i}\frac{\rho_{i}}{s_{i}}
\\ & \geq & \sum_{i=1}^{n}(x_{i}-y_{i})s_{i}\frac{\rho_{j}}{s_{j}}
\\ & \geq & \frac{\rho_{j}}{s_{j}}\sum_{i=1}^{n}(x_{i}-y_{i})s_{i}
\\ &\geq & 0 \ ; \ \text{by Inequality} \ (\ref{eq 15})
\end{eqnarray*}
\noindent Therefore, the total value of solution $X$ is equal or greater than the total value of any other feasible solution, then the solution of the proposed greedy-knapsack algorithm is an optimal solution. $\Box$

\subsection{Ranking Function}
\noindent
As it was described earlier in subsection 3.2, the greedy-knapsack algorithm applies a ranking function as its greedy function to measure the benefit of selecting and scheduling a given bearer and drive the appropriate resource allocation decision. The main idea of the proposed ranking function was inspired by the normalized ranking function in \cite{r28} which is a combination of four individual ranking functions of QoS metrics (delay, loss, queue depth, and priority). Each ranking function outputs a weighted rank value, bounded in $[0, p_{i}]$, and is calculated as follows:
\begin{equation}
rf(v_{i},p_{i})=p_{i}.\tanh(v_{i})
\end{equation}
where $p_{i}$ is the adjustable weight for each QoS metric assigned by the operator and $v_{i}$ is the normalized value of QoS metric $i$ calculated as follows:
\begin{equation}
v_{i}= \frac{measured\ value\ of\ metric\ i}{ QoS\ constraint\ of\ metric\ i}
\end{equation}
In the normalized ranking function the QCI label, indicating the QoS constraints dedicated to bearers, is the main factor that determines the transmission priority of a particular bearer. However, providing fairness and high throughput performance is a challenging issue in case of scheduling strategy unaware of experienced data rate.
Owing to the fact that users in different time slots, sense different qualities of transmission channel, serving users that are in strong channels, leads to a major maximization of system throughput. In other words, since the channel state information is firmly pertinent to the throughput gain, accounting for channel state variations is one of the most effective ways for system throughput maximization. However, achieving efficiency in a spectral domain and traffic fairness are two challenging issues in conflict. Optimization of the channel capacity utilization brings unfair sharing to the terminals with low QCI values staying at the cell-edges. We overcame this conflicting issue by considering a measure of throughput, normalized by the quantity of past data rate, experienced by each user; making this, it is possible to average the resources evenly among the users and consequently provide fairness along with higher overall throughput. As a result, the overall rank for a given bearer will be calculated as follows:

\begin{equation}
\begin{split}
& \sum rf(v_{i},p_{i}),
\\ & \forall i \in \{\textstyle delay, loss, queue depth, priority, throughput\}
\end{split}
\end{equation}

\noindent where specifically, in case of throughput parameter, the Time Domain Proportional Fair metric defined in \cite{r34} will be used as the normalized throughput value. This metric is expressed as:

\begin{equation}
v_{th}=\frac{Wideband Estimated Throughput}{Past Average Throughput}
\end{equation}
The supportable wideband throughput is estimated by the link adaptation, utilizing CQI value and past average throughput, which is the data rate history of each user that is updated every TTI when a bearer is resource allocated.

\section{Simulation}
\noindent
In this section, we explain the simulated environment, traffic model and performance metrics used to evaluate the effectiveness of the greedy-knapsack algorithm. The simulation parameters are listed in Table \ref {table 2}. To evaluate the performance of the proposed optimized resource allocation method, we compare its performance with the traditional priority-only scheme as well as the knapsack algorithm, as the reference algorithms through the same simulation platform applied and stated in \cite{r28}. This simulation environment was implemented based on the LTE network characteristics defined in the 3GPP LTE verification framework \cite{r35} comprising the scheduling aspects of eNodeB MAC layer. There are two different system models, single-cell and multi-cell that can be used for simulation. However, in the wrap-around multi-cell model where the eNodeBs in adjacent cells are assigned channel groups, which are different channels from the neighbouring cells, the first tier and further adjacent cells do not have any influence on the time-domain  scheduling process of the proposed algorithm and the performance measurements for the selected central cell. Therefore, a single-cell scenario is imposed where there is a mixture of different traffic types, as shown in Table \ref {table 3}. The voice and data traffic were modelled by means of exponential distribution function and aggregate self-similar pattern \cite{r36} respectively, to be realistic models, particularly in overload states. These data and voice traffic generators simulate bearers from various QCI classes.
\begin{table*}[htbp]
\caption{Simulation Parameters}
\center
\begin{tabular}{p{7.5cm}ll}

\hline
\hline
Parameter                                 & Value                    \\  \hline
Bandwidth                                 & 5 MHz                    \\
Number of RBs                             & 25 RBs per spectrum allocation (12 subcarrier per RB)\\
Simulation time                           & 32 minutes               \\
Scheduling time window                    & 10 ms                   \\
Number of data bearers                    & 100                      \\
Hurst parameter for data traffic          & 0.9                      \\
Number of voice beares                    & 300                      \\
Voice Activity Factor for voice generator & 0.5                      \\
Mean talk spurt duration                  & 5 s                     \\
Voice codec encoding frame size           & 20 ms                   \\
Modulation and coding scheme              & QPSK, 16QAM and 64QAM    \\
\begin{tabular}[c]{@{}l@{}}Weight of throughput, loss, delay, queue \\ depth and priority metrics respectively\end{tabular}              & 4, 4, 16, 4, 2                        \\
Normal, overload run intervals in sequence    & (1s, 5s, 3s, 2s, 1s, 2s, 1s, 10s, 3s, 5s, 3s) \\  \hline

\end{tabular}
\label{table 2}
\end{table*}

\begin{table}
\caption{Expected application traffic profile \cite{r28}}
\begin{tabular}{lp{1.3cm}ll}
\hline
\hline

Traffic type                     & percentage of users & QCI          \\  \hline
Best effort (FTP)                & 10                  & 6, 8 or 9    \\
Interactive (web browsing /HTTP) & 20                  & 6, 7, 8 or 9 \\
Streaming (video streaming)      & 20                  & 2 or 4       \\
Real time (VoIP)                 & 30                  & 1            \\
Interactive real-time (gaming)  & 20                  & 3            \\  \hline
\end{tabular}
\label{table 3}
\end{table}
In this work, we evaluated the performance of the proposed greedy-knapsack algorithm in terms of the QoS parameters and system average throughput per QoS class. The QoS analysis was made by measuring the correspondent QCI's packet loss rate and packet delay budget. These metrics describe the user's perspective performance and are measured when there are various stochastic intervals of normal and overload states.

\section{Results and Discussion}
\noindent In this research, we seek to solve the knapsack scheduling problem and optimize the total performance of the network by exploiting the greedy property of LTE resource sharing. Furthermore, a QoS and throughput aware ranking function was included to deal with the challenging issue of give-and-take scheduling targets. To better conceive the obtained results, several graphs and tables were generated based on the simulation outputs.

\subsection {Throughput and Fairness}
\begin{table*}
\caption{Average Throughput (Mbps) per QCI class}
\center
\begin{tabular}{p{4cm}lp{1cm}lp{1cm}lp{1cm}lp{1cm}lp{1cm}lp{1cm}lp{1cm}lp{1cm}lp{1cm}}
\hline
\hline

Scheduler           &   QCI1   &   QCI2    &  QCI3    &  QCI4    &  QCI6     &  QCI7    &  QCI8     &  QCI9   \\ \hline
Greedy-Knapsack     &  5.44   &  2.82    &  4.61    &  2.83    &  1.18     &  1.62    &  1.63     &  1.36     \\
\\
Priority            &  5.44   &  2.41    &  5.02    &  2.73    &  1.83     &  1.51    &  1.96     &  0.65     \\
Improved percentage(\%) &  0.00   &  17   &  -8.2 &  4.3  &  -35.5 &  7.3   &  -16.8 &  109.2 \\
\\
Knapsack            &  5.42   & 2.51    &  5.49    & 2 .36    &  1.54     &  1.44    &  1.52     &  1.19     \\
Improved percentage(\%) &  0.4 &  12.4  &  -16   &  19.9  &  -23.4  &  12.5 &  7.2   &  15.5 \\ \hline
\end{tabular}
\label {table 4}
\end{table*}

The average throughput gained by different scheduling approaches, priority only, knapsack and greedy-knapsack, with respect to the various classes of QoS, is shown in Table \ref{table 4}. The greedy-knapsack algorithm shows a general throughput increase especially for QCI classes 1, 2, 4, 7, 8 and 9, compared to the knapsack algorithm. This increment is obtained thanks to the QoS class-based ranking function combined with the time-domain normalized throughput ratio. It turned the multi-service resource allocation algorithm into an opportunistic scheduler, which provides system throughput improvement compared to the QoS guarantee in traffic overload patterns. The greedy-knapsack algorithm tends to give higher priority to the bearers with higher potential of the wideband throughput when they are in superior quality of the channel state. Furthermore, it seeks to give a relatively equal share of the resources to proportionally equalize the throughput of all users. In contrast, the knapsack and Priority Only schedulers have a QoS-oriented allocation pattern, which is completely independent of the frequency domain setting; therefore, the QoS constraints are provided at the expense of the throughput. In the case of the Priority Only algorithm, a bearer having higher QCI priority is resource allocated in advance. Consequently, the biggest throughput rate improvement belongs to the lowest priority QCI class 9, with 109.2\% raise.
There is no significant variation of throughput for QCI class 1 when using the three mentioned algorithms. They show the highest throughput performance around 5 Mbps for VoIP bearers. We can justify this no-variation and maximum throughput because of the prominent factor of QCI metrics, particularly the highest priority affected, in the bearers' sorting pattern.

\begin{figure*}
\center
\includegraphics [width=4.5in,height=2.3in]{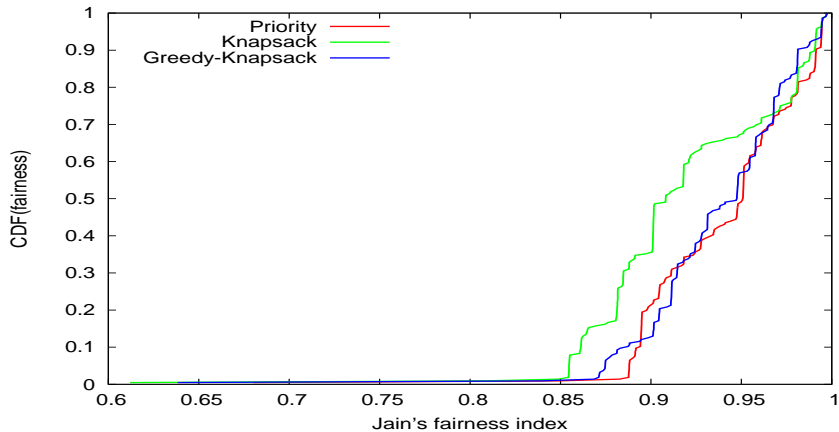}
\caption{Jain's instantaneous fairness index for VoIP bearers}
\label{fig:fair}       
\center
\end{figure*}

To illustrate the relative equal share of the resources, we have evaluated the fairness of the VoIP bearers by computing the Cumulative Distribution Function (CDF) of Jain's fairness index during the simulation window time. As much as the effect of past experienced throughput awareness is more prominent for prioritizing the bearers with the same QoS characteristics, and the VoIP traffic is the major volume of the existing wireless communication traffic the measure of fairness for the VoIP traffic is important in assessing how fair the system scheduling solution performs. Based on Fig. \ref{fig:fair} we can deduce that the level of fairness was improved as the particular effect of the added normalized throughput metric. It indicates that greedy-knapsack achieves a good level of fairness (Jain's fairness index between 0.87 and 1.00) where, 60\% of VoIP bearers receive less than or equal to 0.96 fairness index while for the case of knapsack scheduler in use, they receive less than or equal to 0.92 fairness index.

\subsection{Loss and Delay}
\begin{figure*}
\centering
\subfloat[]{
\includegraphics [width=0.5\textwidth]{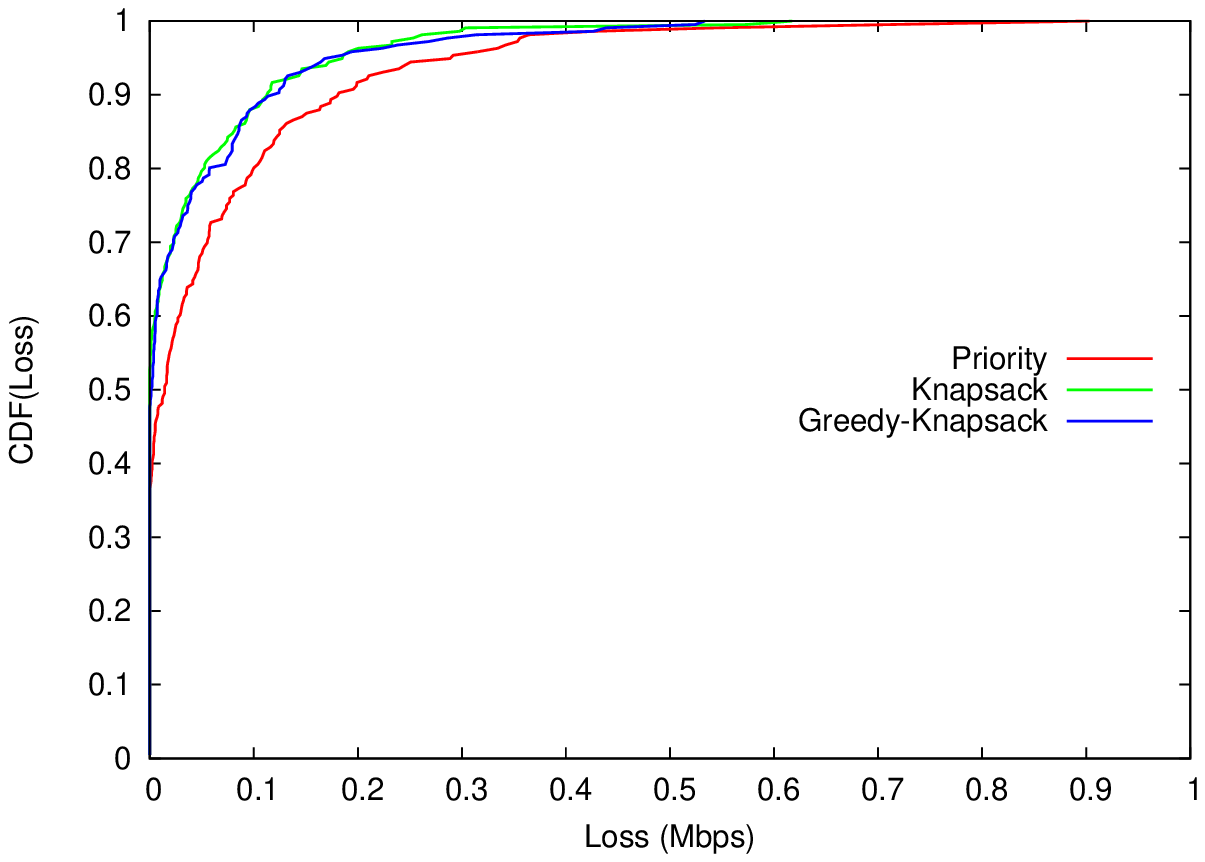}
\label {fig:l2}
}
\subfloat[]{
\includegraphics  [width=0.5\textwidth]{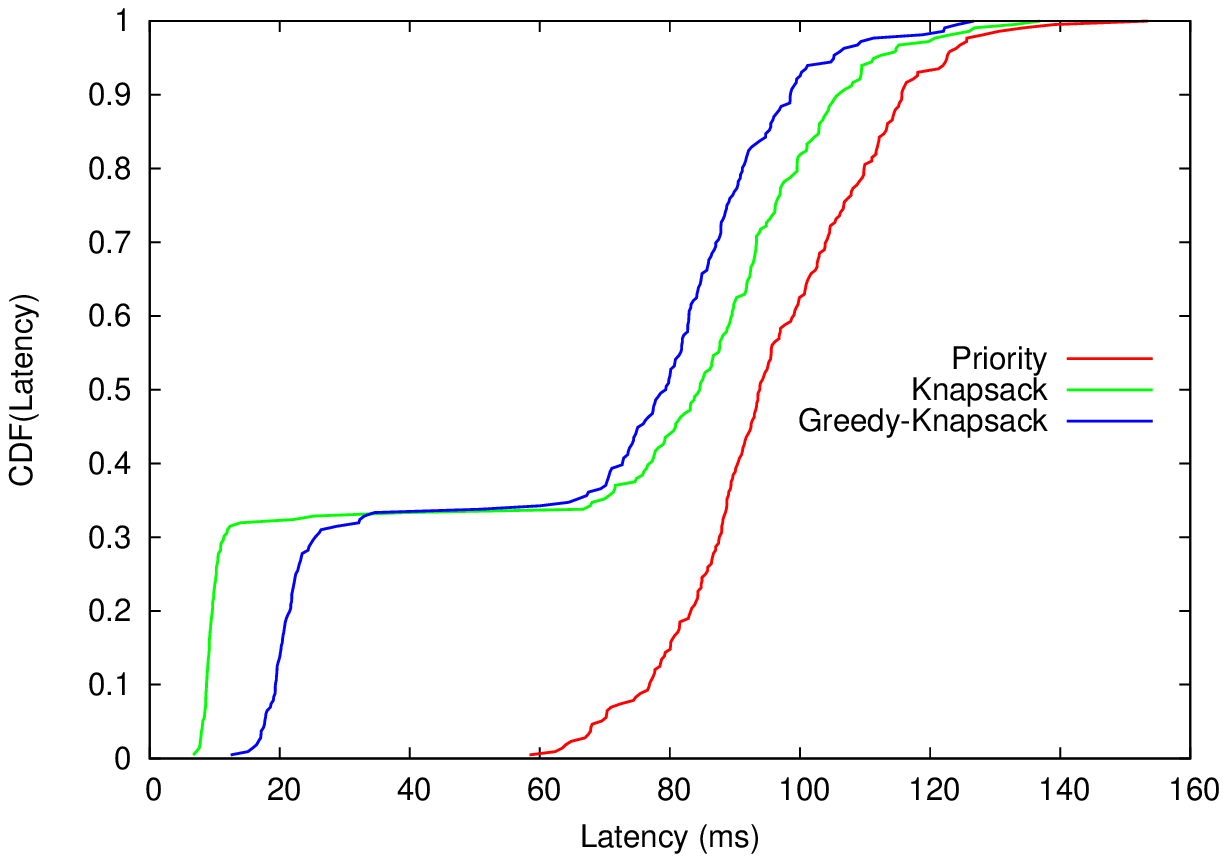}
\label {fig:d2}
 }
\caption{CDF of (a) packet loss and (b) average latency for bearers from QCI class 2}
\label {fig:2}
\end{figure*}

\begin{figure*}
\centering
\subfloat[]{
\includegraphics [width=0.5\textwidth]{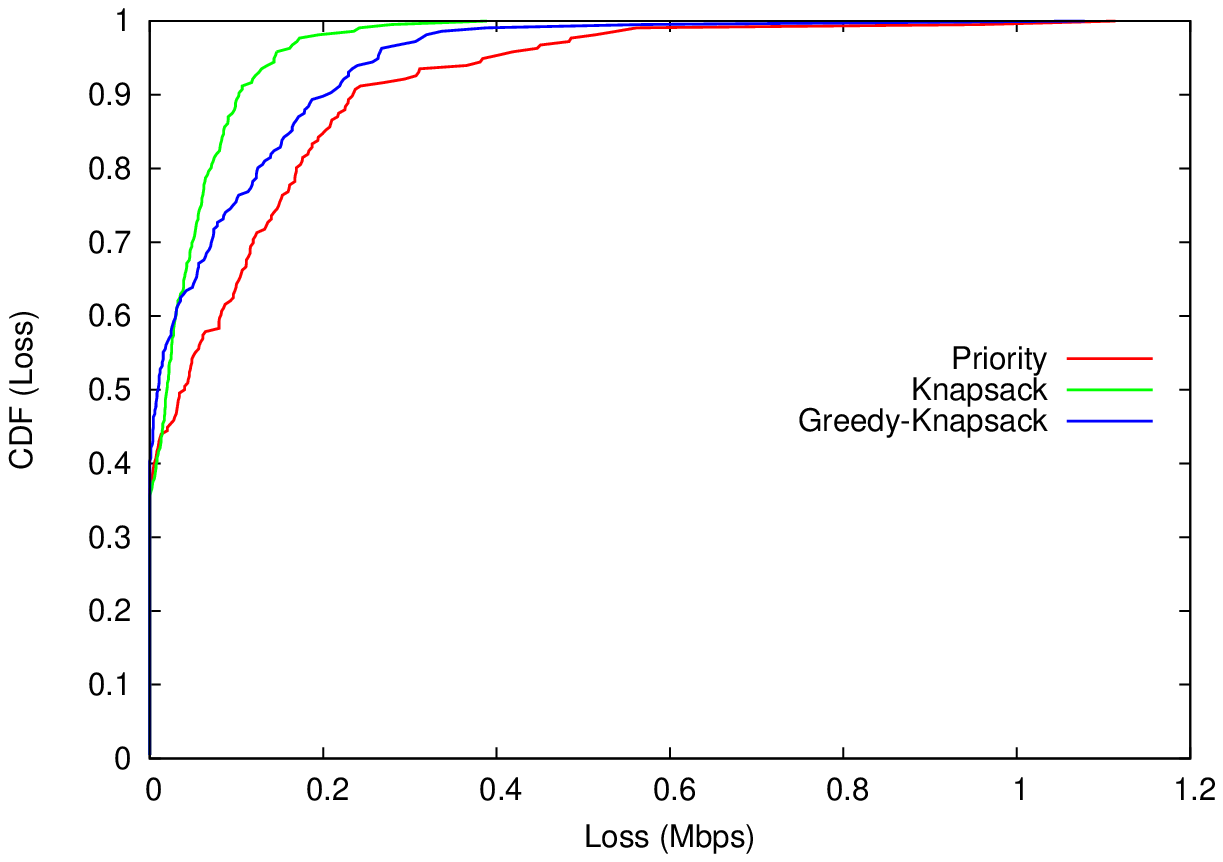}
\label {fig:l3}
}
\subfloat[]{
\includegraphics [width=0.5\textwidth]{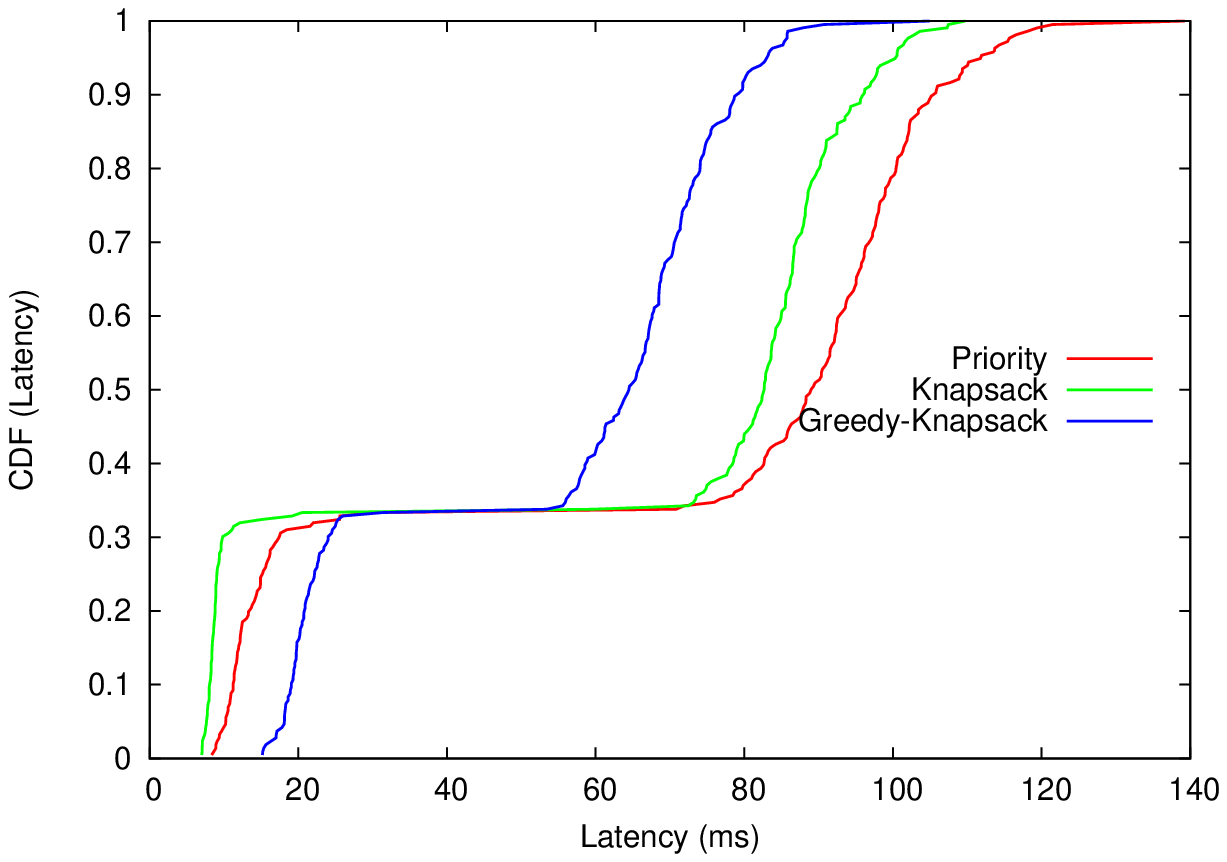}
\label {fig:d3}
 }
\caption{CDF of (a) packet loss and (b) average latency for bearers from QCI class 3}
\label {fig:3}
\end{figure*}

\begin{figure*}
\centering

\subfloat[]{
\includegraphics [width=0.5\textwidth]{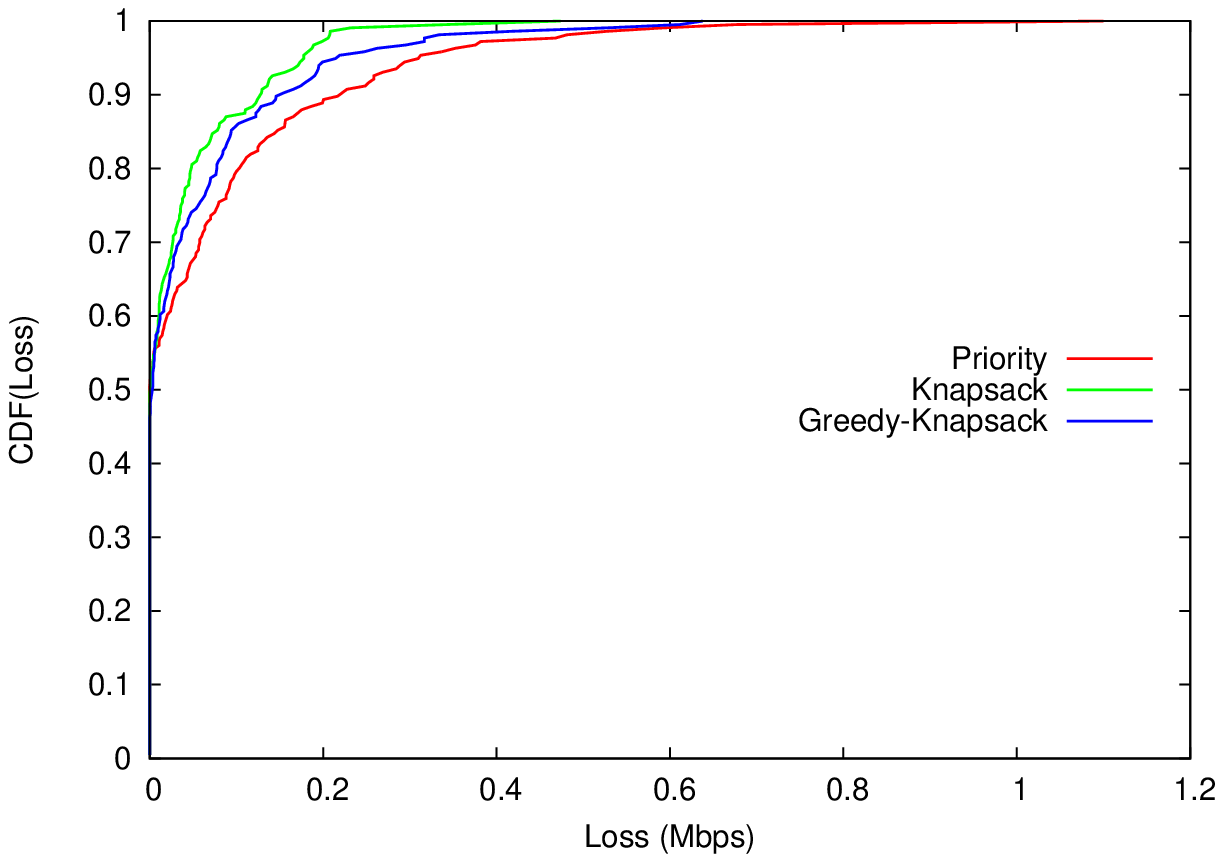}
\label {fig:l4}
}
\subfloat[]{
\includegraphics [width=0.5\textwidth]{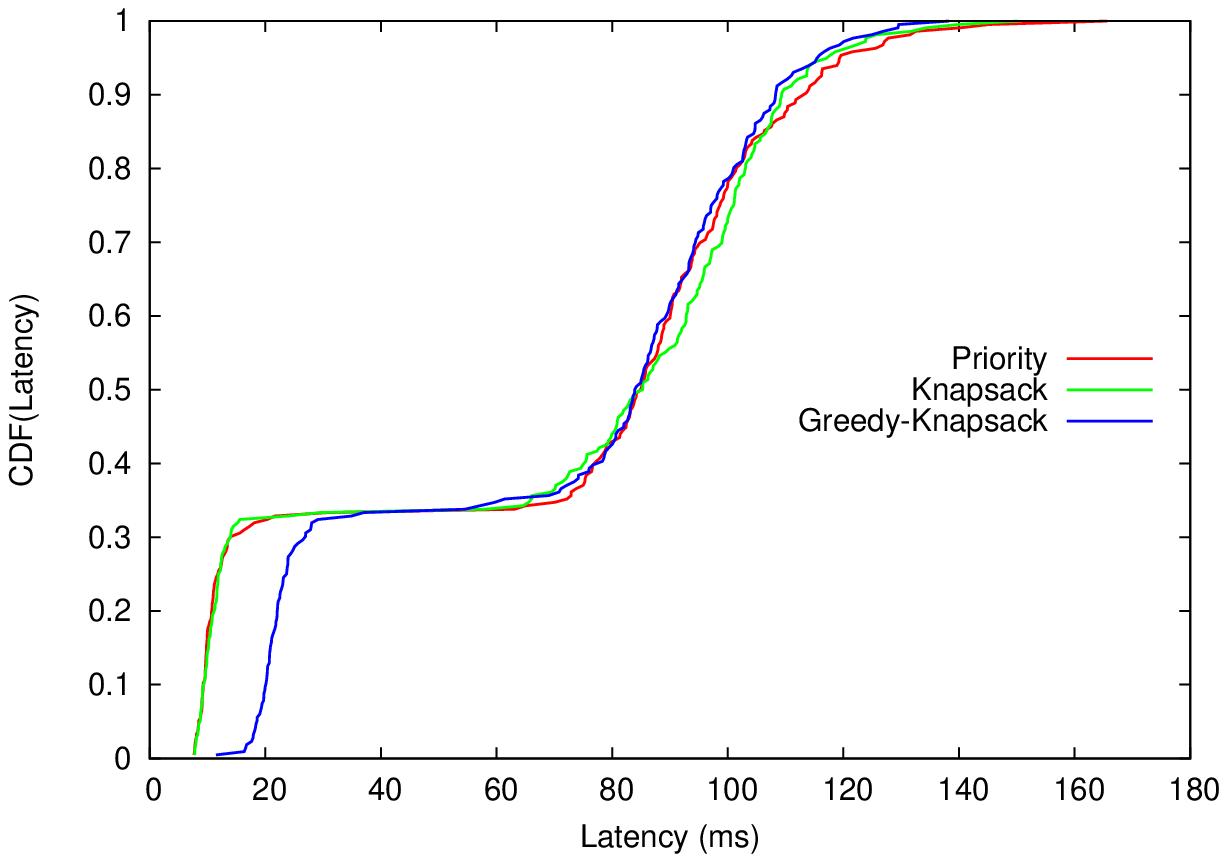}
\label {fig:d4}
 }
\caption{CDF of (a) packet loss and (b) average latency for bearers from QCI class 4}
\label {fig:4}
\end{figure*}

\begin{figure*}
\centering
\subfloat[]{
\includegraphics [width=0.5\textwidth]{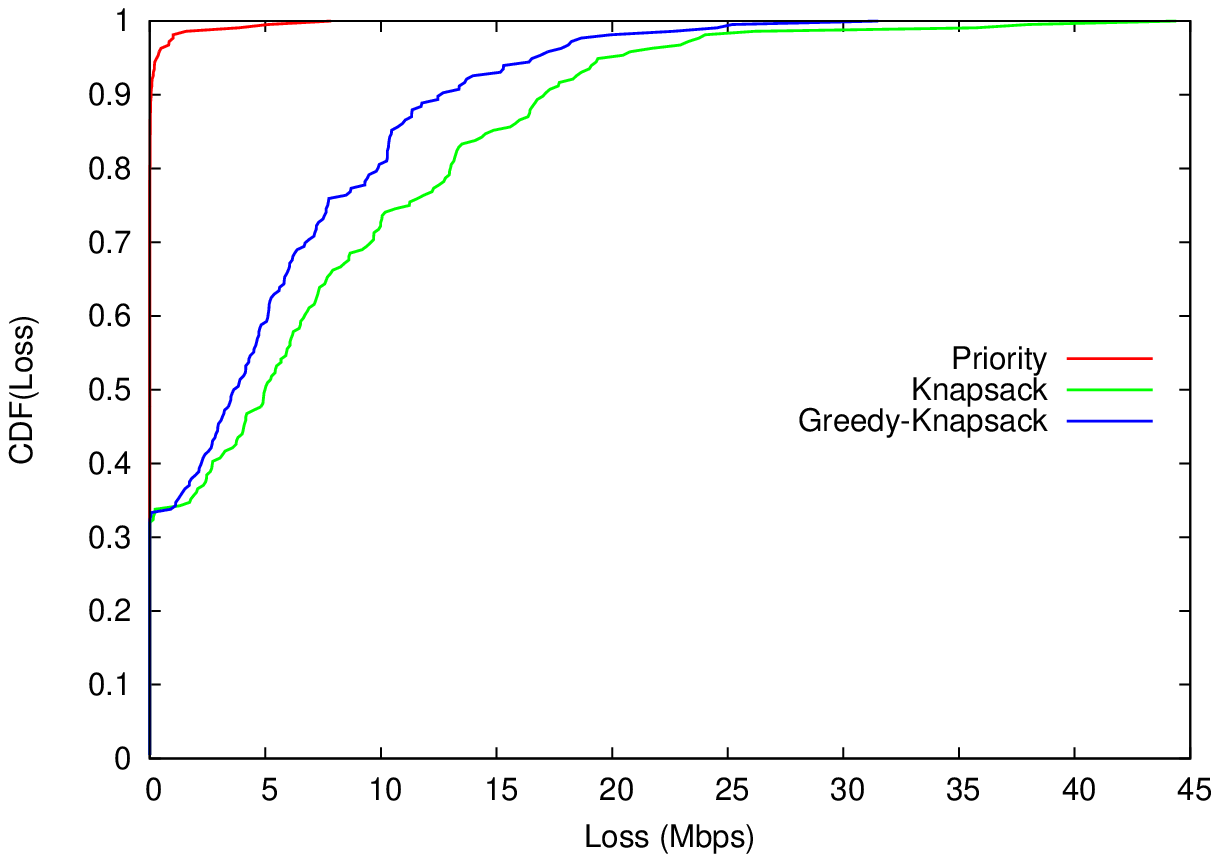}
\label {fig:l6}
}
\subfloat[]{
\includegraphics  [width=0.5\textwidth]{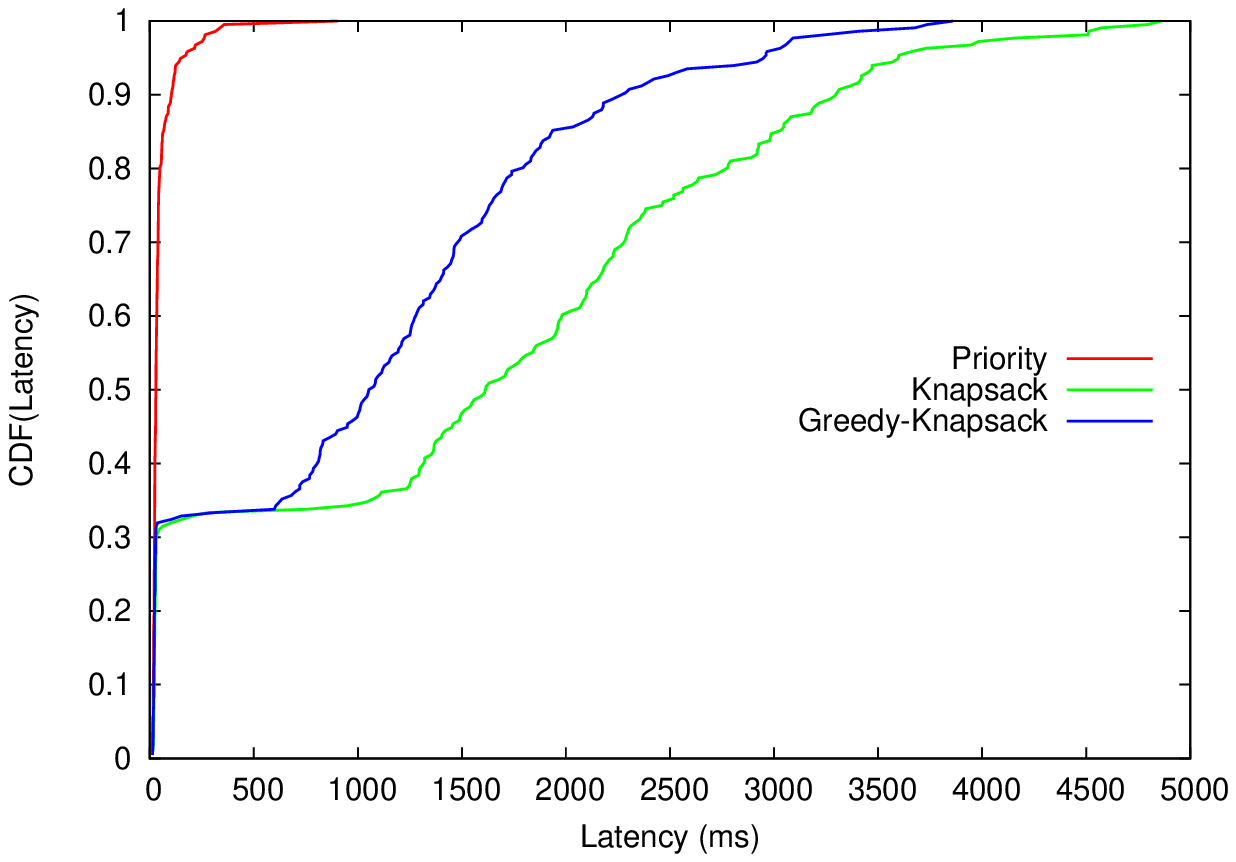}
\label {fig:d6}
 }
\caption{CDF of (a) packet loss and (b) average latency for bearers from QCI class 6}
\label {fig:5}
\end{figure*}

\begin{figure*}
\centering
\subfloat[]{
\includegraphics [width=0.5\textwidth]{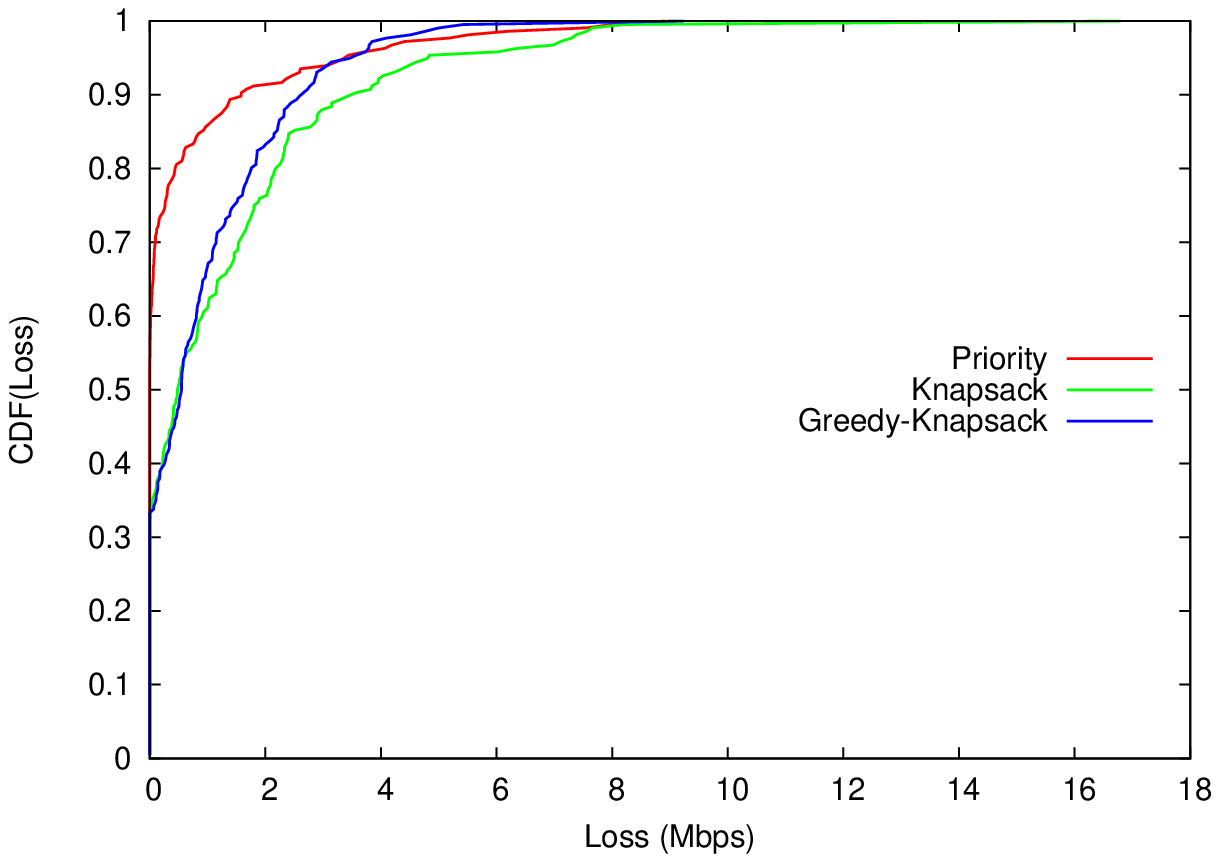}
\label {fig:l7}
}
\subfloat[]{
\includegraphics  [width=0.5\textwidth]{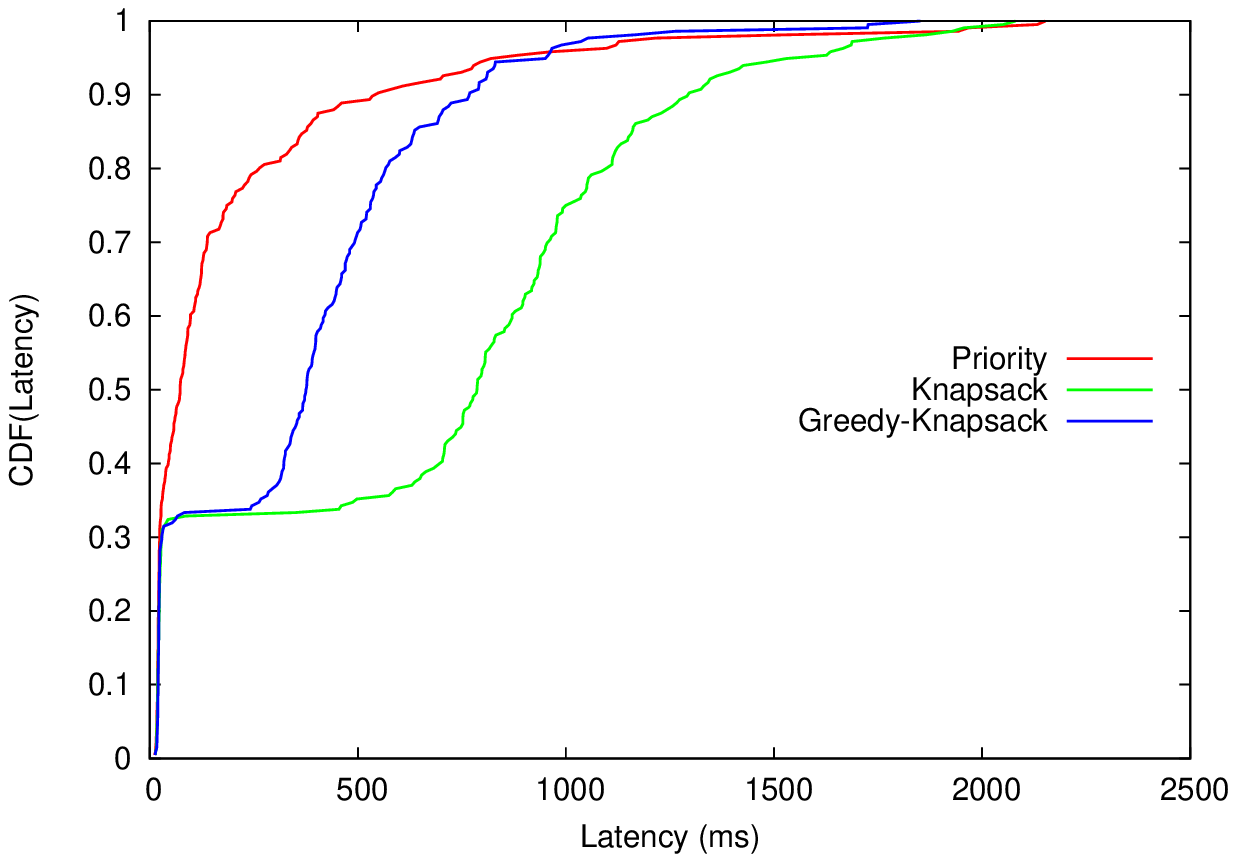}
\label {fig:d7}
 }
\caption{CDF of (a) packet loss and (b) average latency for bearers from QCI class 7}
\label {fig:6}
\end{figure*}

\begin{figure*}
\centering
\subfloat[]{
\includegraphics [width=0.5\textwidth]{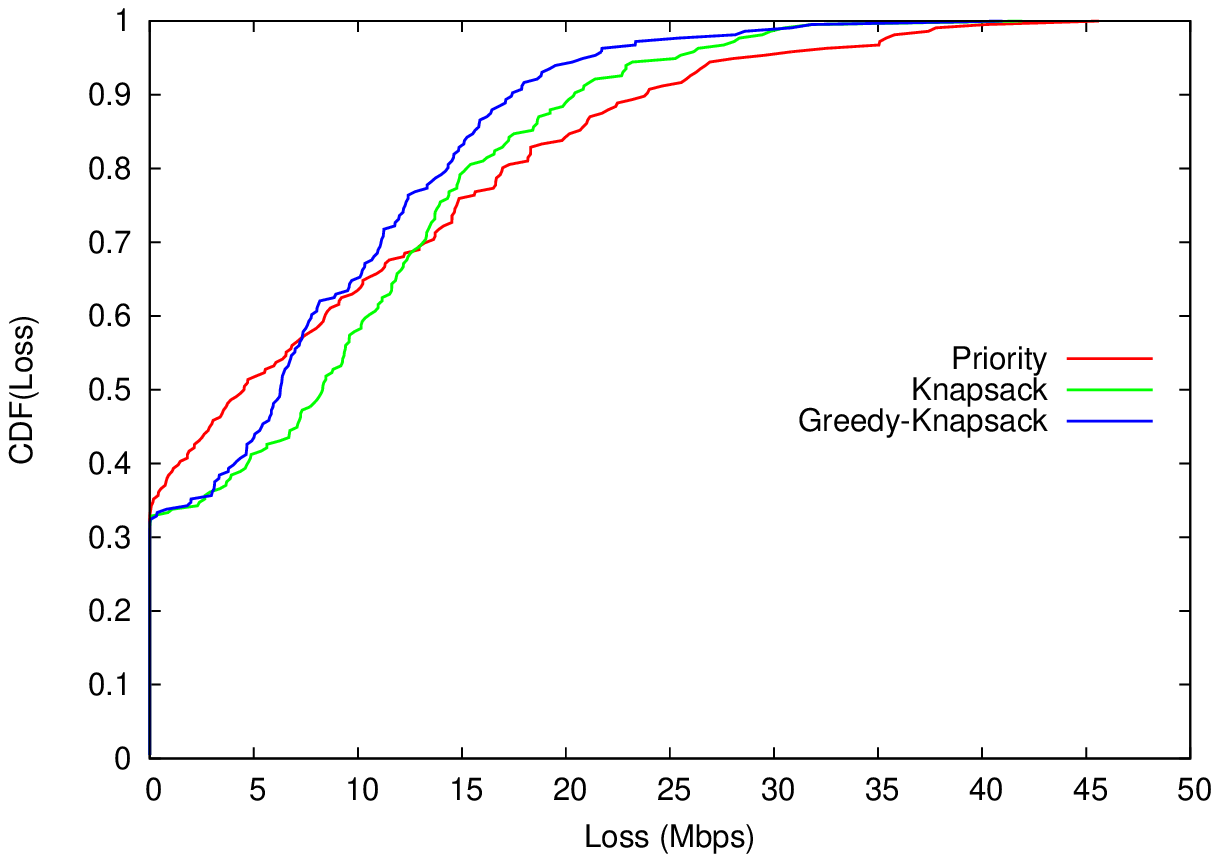}
\label {fig:l8}
}
\subfloat[]{
\includegraphics  [width=0.5\textwidth]{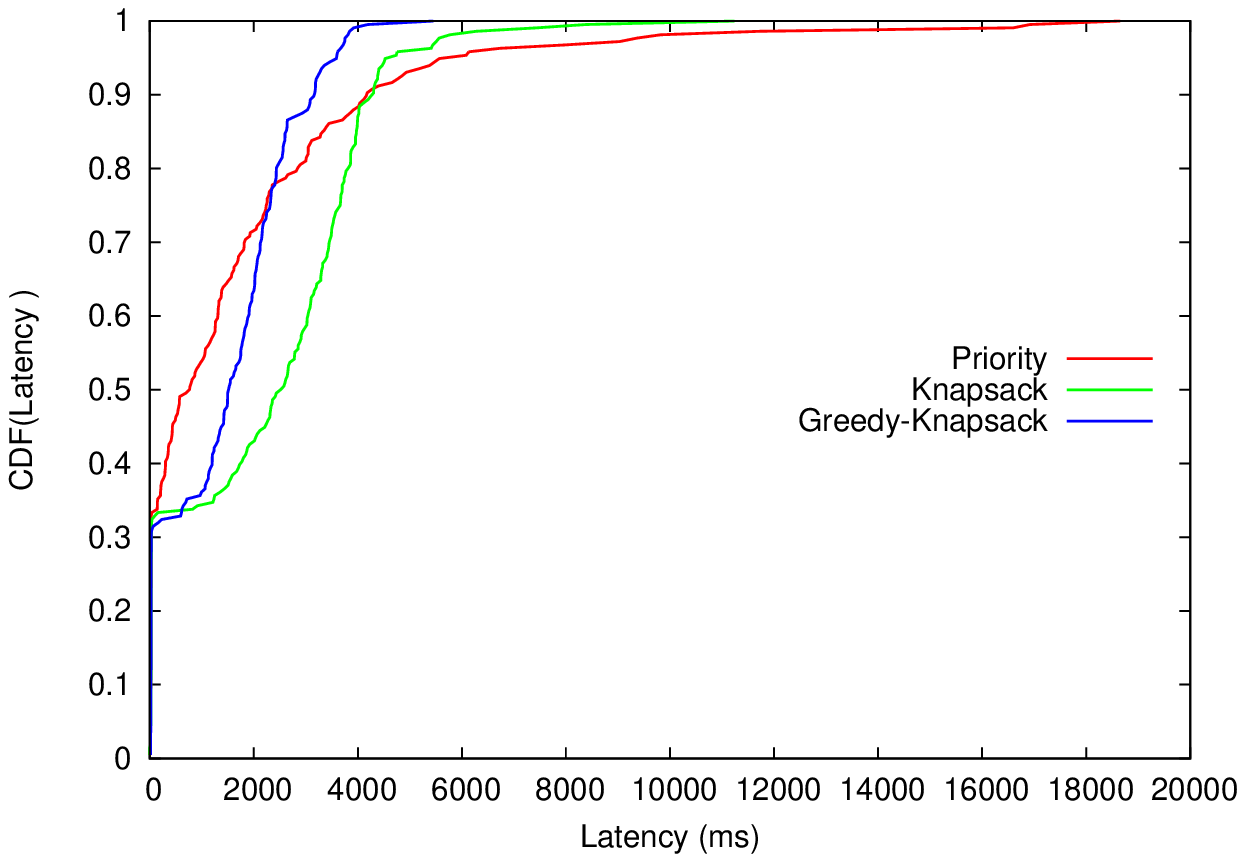}
\label {fig:d8}
 }
\caption{CDF of (a) packet loss and (b) average latency for bearers from QCI class 8}
\label {fig:7}
\end{figure*}

\begin{figure*}
\centering
\subfloat[]{
\includegraphics [width=0.5\textwidth]{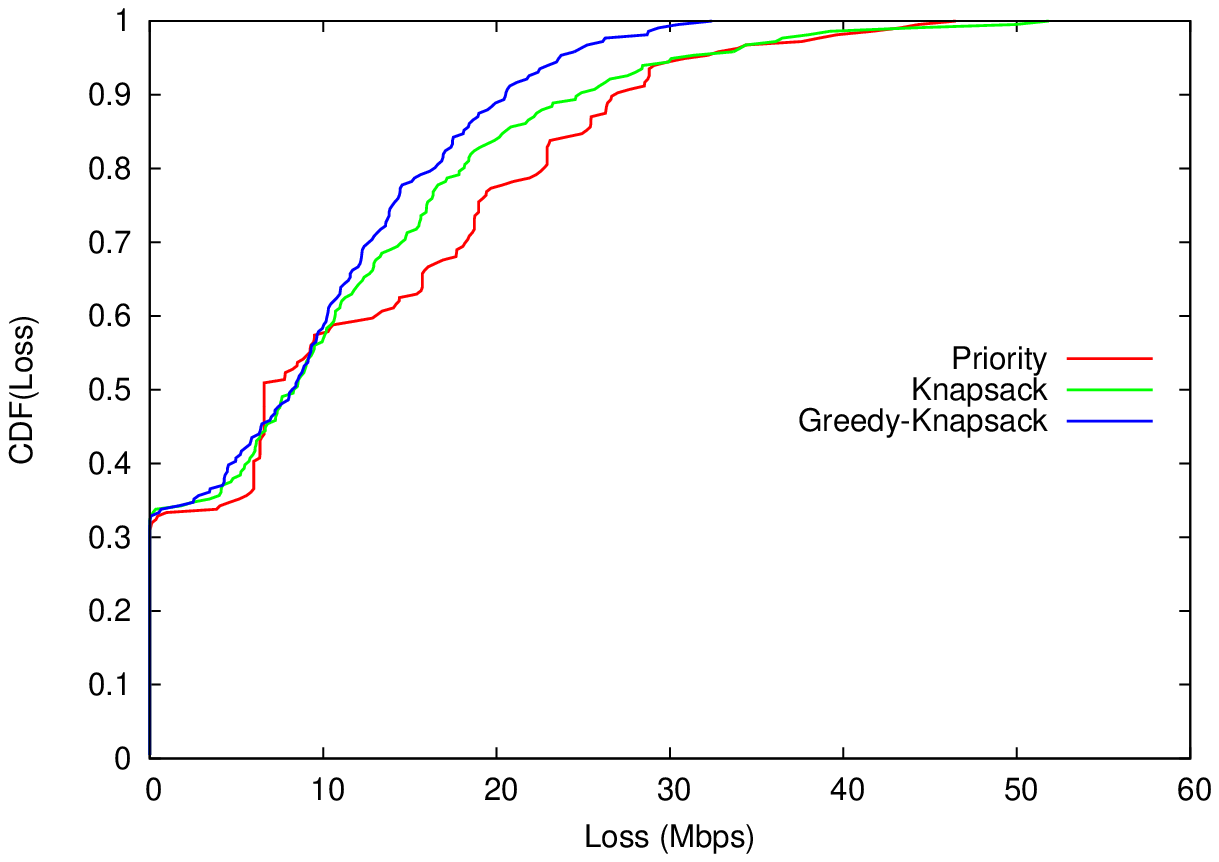}
\label {fig:l9}
}
\subfloat[]{
\includegraphics  [width=0.5\textwidth]{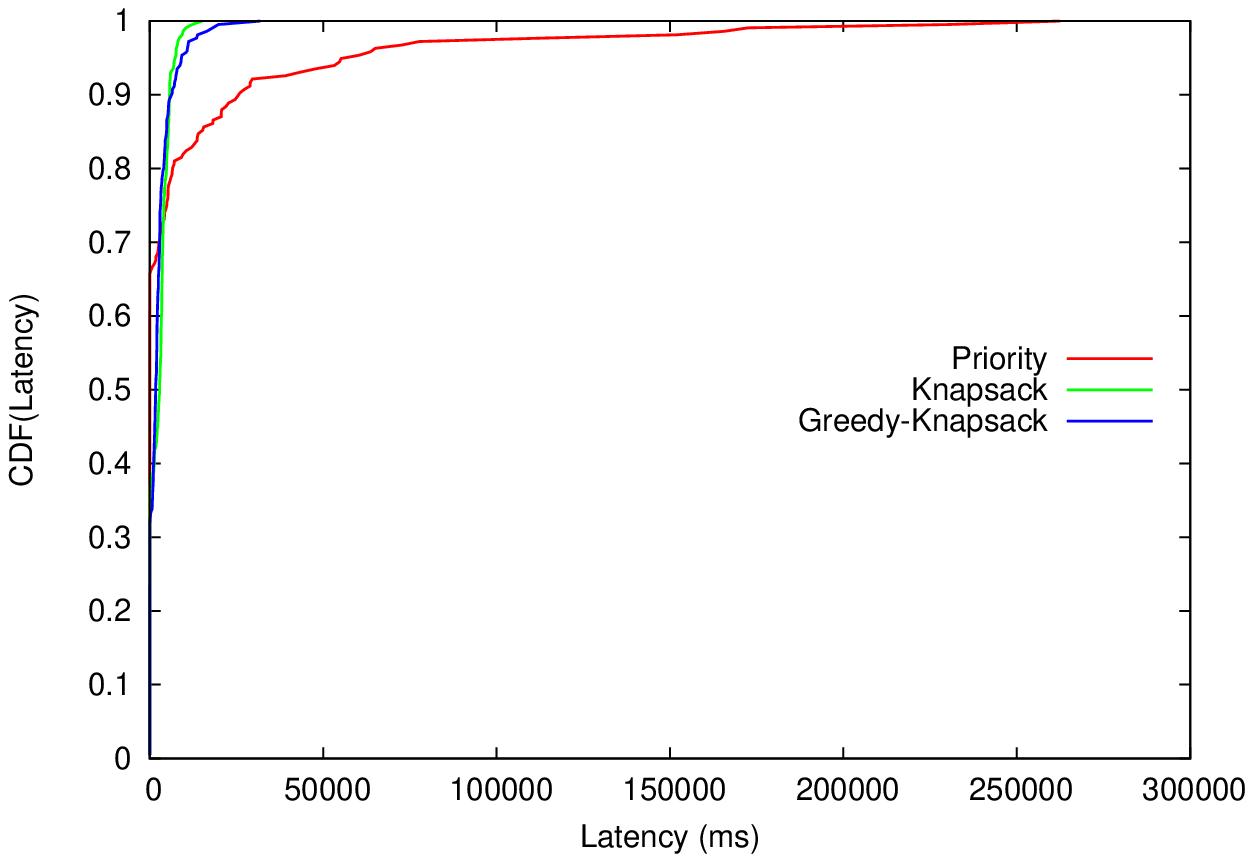}
\label {fig:d9}
 }
\caption{CDF of (a) packet loss and (b) average latency for bearers from QCI class 9}
\label {fig:8}
\end{figure*}

\noindent
To study and compare the behaviour of the greedy-knapsack algorithm with the knapsack and priority only algorithms, Figs. \ref{fig:2}-\ref{fig:8} show the obtained numerical results for the loss and delay, which are defined as the influential QoS factors in providing an optimal level of Quality of Experience (QoE) during the whole simulation time, including intervals of normal and overload states.

Due to the fact that the entire quantity of traffic is greater than the available system bandwidth during the overload periods, the schedulers serve the bearers with the most emergency demand of resource to optimize the performance with a trade-off between the system optimization targets; therefore, no scheduling algorithm would be optimal for all QCI classes in overload periods. Tables \ref{table 5} and \ref{table 6} show the improved percentage of the greedy-knapsack algorithm in terms of average loss and latency per class in comparison with the reference algorithms. As can be seen from these tables, greedy-knapsack algorithm provided a trade-off in terms of loss and delay among the GBR and non-GBR application classes. It has improvement for the most classes especially for QCI 2.

\begin{table*}
\caption{Average Loss (Mbps) per QCI class}
\center
\begin{tabular}{p{4cm}lp{1cm}lp{1cm}lp{1cm}lp{1cm}lp{1cm}lp{1cm}lp{1cm}lp{1cm}lp{1cm}}
\hline
\hline

Scheduler           &   QCI1   &   QCI2    &  QCI3    &  QCI4    &  QCI6     &  QCI7    &  QCI8     &  QCI9   \\ \hline
Greedy-Knapsack     &  0.00   &  0.03   &  0.06    &  0.04    &  5.12     &  0.96    &  7.53     &  8.70     \\
\\
Priority            &  0.00   &  0.06    &  0.10    &  0.07    &  0.12     &  0.53    &  8.74     &  11.36     \\
Improved percentage(\%) &  0.00   &  50   & 40 &  42.9  &  -4,166.6 &  -81.1   &  13.7 &  23.4 \\
\\
Knapsack            &  0.00   & 0.04    &  0.04    & 0.03    &  6.85     &  1.30    &  8.84     &  10.20     \\
Improved percentage(\%) &  0.00 &  25.0  &  -50.0   &  -33.3  &  25.2 &  26.1 &  14.8   &  14.7 \\ \hline
\end{tabular}
\label {table 5}
\end{table*}

\begin{table*}
\caption{Average Latency (ms) per QCI class}
\center
\begin{tabular}{p{4cm}lp{1cm}lp{1cm}lp{1cm}lp{1cm}lp{1cm}lp{1cm}lp{1cm}lp{1cm}lp{1cm}}
\hline
\hline

Scheduler           &   QCI1   &   QCI2    &  QCI3    &  QCI4    &  QCI6     &  QCI7    &  QCI8     &  QCI9   \\ \hline
Greedy-Knapsack     &  4.5   &  65.0   &  53.6    &  70.0    &  1065.4     &  370.8    &  1,458.5    &  2,630.5   \\
\\
Priority            &  4.5   &  95.9    &  68.5    &  67.8    &  51.1    &  203.4    &  1,696.2    &  10,771.3   \\
Improved percentage(\%) &  0.0   &  32.2   &  21.8 &  -3.2  &  -19.8 &  -82.3   &  14.0 &  75.6 \\
\\
Knapsack            &  4.5  & 65.7    &  61.2   & 67.7   &  1,567.5    &  677.6   &  2,229.4    &  2,665.9     \\
Improved percentage(\%) &  0.0 &  1.1  &  12.4   &  -3.4  &  32.0  &  45.3 &  34.6   &  1.3 \\ \hline
\end{tabular}
\label {table 6}
\end{table*}

In response to the guarantee data rate target the GBR bearers are first resource allocated at the expense of non-GBR low priority bearers. Consequently, all three presented algorithms perform strong enough to ensure that the GBR QCI classes 1-4 meet their QoS constraints in terms of loss and delay, resulting in strong QoE for all GBR traffic. The VoIP bearers that correlate to QCI class one are scheduled with no loss, and almost no delay. Conversational traffic from QCI class 2 and the rest GBR bearers from QCI class 3 and 4 experience near to zero loss. Around 90\% of QCI 2 bearers experience loss 0.1 (ms) (Fig. \ref{fig:l2}) and around 90\% of QCI 3 and 4 experience loss less than 0.2 (ms) (Figs. \ref{fig:l3} and \ref{fig:l4}) under the greedy-knapsack and knapsack algorithms. In terms of average latency, greedy-knapsack scheduler has the most improvement for QCI 2 and 3 (Figs. \ref{fig:d2} and \ref{fig:d3}) and almost the same result for QCI 4 latency (Fig. \ref{fig:d4}) in comparison with knapsack and priority only schedulers.

In the case of the non-GBR QCI classes, the difference between the greedy-knapsack scheduler and alternative schedulers increases. As can be seen from Figs. \ref {fig:5}-\ref {fig:7}, the greedy-knapsack algorithm shows a better level of QoE in terms of loss and latency over the QCI classes 6-8 and slightly over QCI class 9 (Fig. \ref{fig:8}), compared to the knapsack algorithm. For example, 90 \% of the bearers from QCI class 6 experience less than 12 (Mbps) loss and 2200 (ms) delay over greedy-knapsack and less than 17(Mbps) loss and 3300 (ms) delay over the knapsack algorithm. This performance improvement is explained because the greedy-knapsack algorithm considers the quantity of packets from each application bearer waiting for scheduling in the bearer's queue.

As shown in Figs. \ref{fig:5} and \ref{fig:6}, the Priority Only algorithm provides a good service with small loss and delay to QCI classes 6 and 7, thus leaving a small quota of bandwidth for QCI classes 8 and 9, which is a weakness of this algorithm. It results in intensely high spikes of delay, especially in case of QCI class 9 in which the bearer waits starving in overload periods and is not served (Fig. \ref {fig:d9}). The low priority bearers of QCI 8 and 9 are the dominant traffic in the current Internet browsing including a huge amount of data flows coming from the applications, such as web pages, email services, peer-to-peer file sharing and progressive video. Regarding this issue, the knapsack and greedy-knapsack algorithms provide an efficient service level, which is finely tuned between all QCI classes 6-9 to handle the bandwidth sharing in the overload conditions of the network.

\section{Conclusion}
\noindent
In this paper, the problem of downlink multi-service scheduling for LTE systems was addressed. In this context, we determined how the candidate users should be selected in each TTI for scheduling, such that the service requirements and bandwidth constraints in the network are fulfilled, without sacrificing the throughput performance of the system. Accordingly, a greedy approach proposed by exploiting the greedy property of the fractional knapsack problem to list an optimal set of users to efficiently share resources by multiple applications. Ultimately, this approach provided an optimal solution to the LTE resource allocation problem, formulated based on the fractional knapsack optimization problem. A throughput-aware class-based ranking function, included in the greedy-knapsack algorithm, was presented to provide a joint optimization of the throughput parameter and QoS constraints to support mixes of GBR and Non-GBR traffic. We compared the system throughput induced by the schedulers, packets delay, and packet loss rate of the bearers coming from different classes of QoS under a mix of normal and overload traffic states. The simulation results showed that the proposed algorithm provided a flexible resource allocation strategy for different classes of QoS with separable constraints, while reducing the loss and delay. Moreover, the experienced data rate awareness of the proposed ranking function resulted in increasing throughput even for Non-GBR classes of applications.
The innovative solutions standardized for LTE-Advanced, such as carrier aggregation, can influence the design of the scheduling algorithms. Therefore, it will be interesting to explore the greedy-knapsack formulation when the modulation and coding scheme constraint, and the component carriers assignment, are also considered to make the resource allocation strategy compatible with LTE-Advanced networks as well.

\subsubsection*{Acknowledgments.} \noindent The authors are thankful to Dr. Michael Brehm and Prof. Dr. Ravi Prakash for their valuable contribution from the University of Texas at Dallas. Also, this work has been supported by the Malaysian Ministry of Education under the Fundamental Research Grant Scheme FRGS/2/2014
/ICT03/UPM/02/3.

\end{document}